\documentclass{nature}
\usepackage{latexsym}
\usepackage{amsthm}
\usepackage{amssymb}
\usepackage{amsmath}
\usepackage[all]{xy}
\usepackage{graphicx}
\usepackage{subfigure}
\usepackage{dcolumn}
\usepackage{bm}
\usepackage{bbm}
\usepackage[usenames,dvipsnames]{xcolor}
\usepackage{amsfonts}
\usepackage{cases}
\usepackage[a4paper,pagebackref=false,colorlinks=true,
linkcolor=blue,citecolor=blue,
pdfauthor={ },
pdftitle={ },
pdfsubject={ },
pdfkeywords={ }]{hyperref}
\newcounter{firstbib}

\begin{document}

\title{Generating Giant and Tunable Nonlinearity in a Macroscopic Mechanical Resonator from Chemical Bonding Force}

\author{Pu Huang$^{1,2\dag}$, Jingwei Zhou$^{1\dag}$, Liang Zhang$^{1}$, Dong Hou$^{1,2}$, Wen Deng$^{1,3}$,Chao Meng$^{1}$,Changkui Duan$^{1}$, Chenyong Ju$^{1,2}$,Xiao Zheng$^{1,2}$, Fei Xue$^{3}$, Jiangfeng Du$^{1,2}$\thanks{e-mail: djf@ustc.edu.cn.\\
\indent\hspace{1mm}$^{\dag}$These authors contributed equally to this work.}}

\maketitle

\begin{affiliations}
\item
National Laboratory for Physics Sciences at the Microscale and Department of Modern Physics, University of Science and Technology of China, Hefei, 230026, China
\item
Synergetic Innovation Center of Quantum Information and Quantum Physics, University of Science and Technology of China, Hefei, 230026, China
\item
High Magnetic Field Laboratory, Chinese Academy of Science, Hefei, 230026, China
\end{affiliations}

\begin{abstract}
Nonlinearity in macroscopic mechanical system plays a crucial role in a wide variety of applications, including signal transduction and processing \cite{Aldridge2005PRL,Badzey2005N,Stambaugh2006PRB,Ono2008APE,Venstra2013NC,Almog2007PRL,Karabalin2011PRL,Braakman2014APL}, synchronization \cite{Shim2007SCI,Zhang2012PRL,Matheny2014PRL}, and building logical devices \cite{Mahboob2008NN,Venstra2010APL,Bagheri2011NN}.
However, it is difficult to generate nonlinearity due to the fact that macroscopic mechanical systems follow the Hooke's law and response linearly to external force, unless strong drive is used.
Here we propose and experimentally realize a record-high nonlinear response in macroscopic mechanical system
by exploring the anharmonicity in deforming a single chemical bond. We then demonstrate the tunability of nonlinear response by precisely controlling the chemical bonding interaction, and realize a cubic elastic constant of \mathversion{bold}$2 \times 10^{18}~{\rm N}/{\rm m^3}$, many orders of magnitude
larger in strength than reported previously\cite{Chan2001PRL,Lee2002PRB,Kozinsky2006APL,Sankey2010NP,Suh2010NL,Karabalin2011PRL,Eichler2011NN}.
This enables us to observe vibrational bistate transitions of the resonator driven by the weak Brownian thermal noise at 6~K.
This method can be flexibly applied to a variety of mechanical systems to
improve nonlinear responses, and can be used, with further improvements, to explore macroscopic quantum mechanics \cite{Yurke1986PRL,Peano2004PRL,Katz2007PRL,Savel¡¯ev2006NJP,Serban2007PRL,Sillanpaa2011PRB}.
\end{abstract}
\maketitle


Nonlinearity dynamics of a macroscopic mechanical resonator can be modelled
by a Duffing oscillator. Its equation of motion under external drive can be expressed as:
\begin{align}
\label{Duffing}m \ddot{x}+\gamma \dot{x}+k x+\alpha x^3=F_{\rm drive} {\cos}(\omega t).
\end{align}
Here $m$, $\gamma$ and $k$ are the mass, dissipation rate and linear spring constant of the resonator,
and $\alpha x^3$ is the Duffing nonlinearity with $\alpha$ the Duffing constant (cubic elastic constant). Under weak drive, the nonlinear response is negligible due to its cubic dependence of amplitude $x$ of the resonator, and the resonator behaves
like a simple harmonic oscillator. This is the well-known Hooke's law of elasticity \cite {Landau1986}. On the other
hand, under strong enough drive, nontrivial dynamics of the resonator emerges. A famous example is the existence of Euler instability
where bifurcation occurs when the drive strength reaches a certain threshold $F_c$.
Since the intrinsic Duffing constant in macroscopic mechanical systems is weak \cite{Lifshitz2008}, and as $F_{c}$ is reversely proportional to $\sqrt{\alpha}$,
 very strong drive is needed in order to explore the Duffing nonlinearity for applications \cite{Landau1986}. This brings in various unfavorable factors: it leads to high power consumption \cite{Mahboob2008NN,Venstra2010APL}, and moreover, strong drive brings in strong
parametric noise far beyond the Brownian thermal noise even at room temperature \cite{Aldridge2005PRL,Badzey2005N,Stambaugh2006PRB,Ono2008APE,Venstra2013NC,Karabalin2011PRL,Almog2007PRL,Braakman2014APL,Shim2007SCI,Zhang2012PRL,Matheny2014PRL}. A method to enhance substantially the Duffing constant $\alpha$ for a macroscopic systems beyond its intrinsic limit is therefore of paramount importance.

Our system consists of a macroscopic mechanical resonator tightened to an anchor via a chemical bonding structure shown schematically in Fig.~1a.
Strong nonlinearity is achieved when the resonator moves in the $x$-direction, deforming the chemical bond. This is because, although the resonator alone follows the elasticity theory with linear dynamics, the response of the chemical bond is highly anharmonic. This is confirmed by  the chemical bond's energy curve $U_{\rm chem}(x)$ (Fig.~1b) obtained by density functional theory calculations (see Methods for details).
For simplicity, we have set for the minimum of $U_{\rm chem}(x)$ (Fig.~1b) as $x=0$. By applying on the chemical bond an external control force $F$, the resonator's equilibrium position can be shifted, and the spring constant will be modified by $\Delta k = -\partial^2 U_{\rm chem} / \partial x^2 $. As shown in Fig.~1c, when the resonator's equilibrium position is far away from the atom contact, the chemical bonding interaction is weak. When we tune the control force, i.e., shift the resonator's equilibrium position towards $x=0$, the strength of $\Delta k$ first reaches a local maximum and then drops to zero at the place where the chemical bonding attraction reaches maximum. The Duffing constant from chemical bonding, i.e., $\alpha = -(1/6)\partial^4 U_{\rm chem}/ \partial x^4 $, changes differently (Fig.~1d). Its strength reaches a local maximum, then drops to zero before approaching the maximum attraction point, and finally changes its sign and increases in strength. At the maximum attraction point, $\alpha = 2\times 10^{22}~{\rm N/m^3}$, while $\Delta k$ is zero.

Fig.~1e shows the threshold driving force $F_{c}$ as a function of the macroscopicity of the resonator (here characterized by the resonator's mass) with/without using the chemical bonding force induced nonlinearity. By introducing chemical bonding force at the maximum attraction point, $F_{c}$ can be reduced significantly. The term proportional to $x^3$ in the expansion of chemical bonding potential $U_{\rm chem}(x)$ will generate a d.c. force as well as a second harmonic term of the drive signal, but they contribute nothing to bistate dynamics of our current interests. While higher-order terms beyond the Duffing nonlinearity are negligible.

We demonstrate the above idea by using a macroscopic doubly clamped beam whose fundamental vibrational mode couples to a gold-atom contact, as shown in Fig.~2a. The atom contact is made by electric current migrations (see Methods) on a nano-bridge which anchors the beam to a stiff electrode (Fig.~2b). The dimensions of the beam is $l\times w \times t = 50~{\rm \mu m} \times 1.5~{\rm \mu m} \times 0.51~{\rm \mu m}$. The mass of the beam is approximately $0.2~{\rm ng}$. The vibrations of the beam  deform the gold-gold bonds of the contact
along $x$ direction. The beam has an intrinsic frequency $\omega_{0}/2\pi =1.58~{\rm MHz}$, a quality factor $Q=3100$, and an intrinsic spring constant $k_0=10~{\rm N/m}$. The device  is placed in an ultrahigh vacuum  chamber and has an environment temperature 6~K.


Fig.~3a plots the measured resonator's frequency shift $\Delta \omega /2\pi$ as well as the change of spring constant $\Delta k$
by tuning the control force $F$. The tuning is done by the Lorentz force generated by passing a current through the beam. The resonance frequency of the beam is measured under weak drive, where the response of the beam is in linear regime. From the change of spring constant, we estimate the Duffing constant as
\begin{equation}
\label{alphak} \alpha = \frac{1}{6 \xi^2}\left (\frac{\partial^2 \Delta k}{\partial F^2}  k^2+  (\frac{\partial \Delta k}{\partial F})^2  k\right ),
\end{equation}
with $\xi\approx 0.83$ the shape constant of beam. In order to reliably obtain the Duffing constant from the data with noise, we have smoothed the $\alpha$ using a running average of 5 points. Fig.~3b plots $\alpha$ as a function of the change of spring constant $\Delta k$.

Fig.~3b plots the measured Duffing constant $\alpha$ as a function of $\Delta k$.
The maximum strength of the Duffing constant achieved in our experiments is $(2.1\pm0.3)\times 10^{18} ~{\rm N/m^3}$, with an enhancement of 4 orders compared to the intrinsic nonlinearity due to the elongation of the beam ($\alpha_0= 2.2\times 10^{14}~{\rm N/m^3}$), where the strength of $\Delta k$ is still very small, i.e., only a few percentage of the intrinsic spring constant. When we further shift the beam closer to the electrode, the beam and electrode jump to contact, resulting disappearance of the vibration mode. Such instability can be avoided, for example, by exploring a resonator with larger spring constant or by lowering the environment temperature (see Supplementary Information). In a practical system, apart from the short range chemical bonding force, the Duffing constant can also have contributions from van der Waals (vdW) force of the longer range, whose strength is however, small as shown in Fig.~3b.

We then set the control force at point $c$ in Fig.~3b, and increase the driving strength to observe the nonlinear response of the beam. The observed hysteresis response is plotted in Fig.~3c. It is different from the hysteresis response of intrinsic nonlinearity due to the elongation of the beam, as shown in Fig.~3d. The observed hysteresis response in Fig.~3c shows an opposite direction, and the corresponding driving force is much weaker (about two orders smaller), indicating a negative and much stronger Duffing constant of the beam, as predicted by the theory \cite{Lifshitz2008}.


In the following we present a demonstration of a dynamical effect of the  strong nonlinear response.
It is known that noise can induce bistate transitions in a driven Duffing oscillator \cite{Dykman1980PA}. However, such transitions in macroscopic mechanical resonators are only observed by introducing strong artificial noise that is much large than the Brownian thermal noise at room temperature, due to weak intrinsic nonlinear response \cite{Aldridge2005PRL,Badzey2005N,Stambaugh2006PRB,Almog2007PRL,Ono2008APE,Venstra2013NC}.
We are able to observe in our system the bistate transitions activated by the Brownian thermal noise even at cryogenic temperatures. In doing so, we tune the control force to a point with the Duffing constant approximately $-1\times10^{17}~{\rm N/m^3}$ and drive the system to bistable state regime. By recording the amplitude of vibrations of the beam we observed a switching behavior, as shown by the green trajectory in Fig 4a.

To verify that such switchings are indeed from the Brownian thermal noise of the beam,
we introduce an amplitude modulation to the driving signal with amplitude as
$F_{\rm drive}(t)=F_{\rm drive}+\delta F_{\rm drive} \cos(\Omega t)$
,with $\Omega=0.5~{\rm Hz}$ and $\delta F_{\rm drive} = 0.18~{\rm pN}$.
The amplitude of vibrations of the beam shows periodic switchings
(Fig.~4a, purple trajectory) instead of random switchings. Fig.~3b shown the corresponding power density,
$S_{\rm mod}(\Omega)$ with modulations and $S_{\rm noise}(\Omega)$ without modulations, from which we define a signal to noise ratio:
${\rm SNR}(\Omega)=S_{\rm mod}(\Omega)/S_{\rm noise}(\Omega)$. Then, by using standard stochastic resonance theory (see Methods), we estimate the total force noise as $\sqrt{S_{\rm total}^{\rm F}}=(3.6 \pm 0.6 )\times 10^{-16}~{\rm N}/\sqrt{\rm Hz}$. It agrees nicely with the resonator's Brownian thermal noise, estimated by $\sqrt{S_{\rm th}^{\rm F}}=4m\omega_{0}k_{B}T/Q$ \cite{PuHuang2013PRL} which equals to $3.3\times 10^{-16}~{\rm N}/\sqrt{\rm Hz}$. In the limit $\Omega $ approaching zero, the modulation becomes a perturbative force signal $ \delta F_{\rm drive} {\rm cos}(\omega t)$ added to the driving force with the same phase. In the rotating frame of the driving signal, the bistate dynamics is described by an over-damped double well whose shape is tuned by $ \delta F_{\rm drive} $ \cite{Dykman1980PA}, and becomes sensitive near the bifurcation point.
By applying a weak force $\delta F_{\rm drive} = 2~{\rm fN}$ that is only several times
larger than the resonator's Brownian thermal noise force, significantly change in the statistics distribution is observed, shown in Fig.~4c. The sensitivity of such response to external force is limited by the total force noise $\sqrt{S_{\rm total}^{\rm F}}$.
In our device, besides the thermal noise, parametric noise from the circuit also contributes to the total force noise, with the weak power density being $\label{Spar} S_{\rm para}^{\rm F} = S_{\rm total}^{\rm F}-S_{\rm th}^{\rm F}$.
We estimate that $S_{\rm para}^{\rm F}$ is $11~{\rm dB}$ below the $S_{\rm th}^{\rm F}$ level, corresponds to a
parametric noise temperature of $500~{\rm mK}$, and is mainly limited by our room temperature detection circuit (see Methods).


In conclusion, we have demonstrated a highly controllable nonlinearity in a macroscopic mechanical system, with its fluctuation dynamics dominated by Brownian thermal noise. The universal existence and the small scale of the chemical bonding force make our method applicable to current widely-used micro- and nano-mechanical systems in improving their nonlinear responses.
With further improvements, nonlinearity induced quantum behaviors in macroscopic mechanics \cite{Yurke1986PRL,Peano2004PRL,Katz2007PRL,Savel¡¯ev2006NJP,Serban2007PRL,Sillanpaa2011PRB} are foreseeable in the type of systems described here.

\begin{methods}
\textbf{Theoretical description of the system.}
Density functional theory calculations were employed to estimate qualitatively the chemical bonding force. Due to stability considerations, a small structure of two gold-atom clusters was adopted in the calculation (see Sec.~1 of Supplementary Information).  By changing only the relative position $x$ of the left cluster relative to the right one while otherwise keeping the relative positions of all the gold atoms fixed, we obtained a chemical bonding energy function $U_{\rm chem} (x)$, from which we extracted the nonlinearity by calculating the fourth order derivative of the total energy. Due to the existing systematic error in performing density functional theory calculations, the second order derivative of the total energy shows slightly an oscillatory behavior at large distance range. We smoothed the calculated data
by fitting it to the function $\frac{A}{x^a}+\frac{B}{x^b}$. In practical case, however,  the relaxation of gold atoms can occur which will change
the energy - displacement $x$ curve. We have considered such effects and found out that the results do not change significantly (see Sec.~1 of Supplementary Information for detailed data and comparison). To estimate the long-range forces, we considered analytically the van der Waals (vdW) attraction and the electrostatic force  \cite{Giessibl2003RMP} for
the geometric model of two gold cylinders close to each other, similar to the case of our nano-bridge structure. To estimate the intrinsic macroscopic mechanical property, we modeled the macroscopic resonator
as a doubly clamped beam. The lowest vibrational mode has a weak intrinsic nonlinearity due to elongation.
In calculating the threshold drive Force $F_c$ in Fig 1e, we took the mechanical quality factor 3000 and the ratio between the beam's thickness and length $t/l \ge 0.005$.

\textrm{\\}
\textbf{Fabrication of the sample.} The device was fabricated on a commercial Silicon-On-Insulator wafer using nanolithography. The nano-bridge connecting the beam and the stiff electrode is about 200 nm in length and 80~nm in diameter. Focus-ion beam (FIB) was adopted to narrow it down to less than 50 nm.
Once such a device, i.e., the doubly clamped beam with a suspended nano-bridge, was successfully
fabricated, we placed it in the ultrahigh vacuum environment of the cryogenic system, and then electro-migrate the bridge to form an atomic point contact  \cite{Park1999APL}. After the atom contact was produced, the beam was disconnected from the stiff electrode and its lowest vibrational mode was measured.
The position of the equilibrium of the beam is controlled by applying a d.c. current through the beam so the atom interaction can be tuned.
The separation of the atom contact can be controlled with a precision about 1~pm without any feedback control for thousands of seconds (see Sec.~2.4 of Supplementary Information for the data). We applied a voltage bias less than 200~mV  between the beam and the stiff electrode to compensate the contact potential of the atom contact so that the contribution of electrostatic force in our experiments can be negligible (see Sec.~3 of Supplementary Information). We decreased the separation of the atom contact gradually by increasing the control force, while monitoring the frequency of the resonator. In doing this, we make the beam vibration amplitude around 50~pm until a jump to contact happens, and after that, a finite resistance (typically around 1~${\rm K \Omega}$) can be detected between atom contact which shows good contact without contamination. Then we decrease the control force until the beam detaches from the stiff electrode. The typical force required to detach the beam is several nN, indicating that the atom contact is a small gold-atom cluster \cite{Rubio-Bollinger2001PRL}.

\textrm{\\}
\textbf{Device characterization and electrical parametric noise.} The Brownian thermal noise on the beam is
$3.3 \times 10^{-16}~{\rm N/}\sqrt{\rm Hz}$, and the thermal motion amplitude noise on resonance is about $1.0\times 10^{-13}~{\rm m}/\sqrt{\rm Hz}$ accordingly. The intrinsic nonlinearity is measured from the frequency response to the driving strength in a standard way based on the Duffing nonlinear model  \cite{Lifshitz2008}.

To do this, we have pulled the beam far apart from the stiff electrode with contact separation larger than 30~nm, so the interaction due to the atom contact is negligible. The intrinsic Duffing constant obtained is $2.2\times 10^{14}~{\rm N/m^3}$ (see Sec.~2.5 of Supplementary Information for data). To estimate the electrical parametric noise in our system, we model the mechanical resonator
as LCR elements  \cite{Cleland21999SAA}, and analyze the noise in the electric circuit (see Sec.~3 in Supplementary Information). The electrical noise
in the circuit produces a current noise $I_{\rm noise}$ passing through the beam, which results in a force noise $ BlI_{\rm noise}$ of power density $S_{\rm para}^{\rm F}$.

 In our experiment, there are mainly three sources of electrical noise.
The first one is the Johnson-Nyquist noise $S_{\rm R}^{\rm V}=4Rk_BT$, with $T$ the room temperature, which produces a force noise with power density $S_{\rm R}^{\rm F}$. The second one is the current leakage from the input of the voltage preamplifier $S_{\rm ba}^{\rm I}$. We model
the preamplifier back-action by a current source similar to that described in reference  \cite{Devoret2000NAT}.
In doing this, we assume that there is no correlation between voltage imprecision $S_{\rm im}^{\rm V}$
and the back-action $S_{\rm ba}^{\rm I}$  \cite{Clerk2010RMP}, so $S_{\rm ba}^{\rm I}$ leads to force noise with power density $S_{\rm ba}^{\rm F}$.
Another one is from the r.f. driving signal, which generates a phase noise and works as a equivalent force noise on the beam with power density $S_{\rm pha}^{\rm F}$.
We estimate the total parametric force noise as: $S_{\rm para}^{\rm F}=S_{\rm R}^{\rm F}+S_{\rm ba}^{\rm F}+S_{\rm pha}^{\rm F}$,
and in the experiments, $\sqrt{S_{\rm para}^{\rm F}}= 0.9 \times  10^{-16}~{\rm N}/\sqrt{\rm Hz}$,
with power density about 11~dB below the $S_{\rm th}^{\rm F}$ (see Sec.~3.2 of Supplementary Information for data).

\textrm{\\}
\textbf{Measurement of total force noise on the beam.}
The dynamics of our system are modeled by the standard Duffing oscillator of equation of motion
\begin{align}
\label{Duffing}m\ddot{x}+\gamma \dot{x}+k x+\alpha x^3=F_{\rm drive}(t)\cos(\omega t)+ F_{\rm noise}(t),
\end{align}
where the dissipation rate $\gamma=m \omega_{0}/Q$, the spring constant $k=m\omega_0^2$, $F_{\rm drive}(t)=F_{\rm drive}+\delta F_{\rm drive} \cos(\Omega t)$ is the amplitude modulated driving force, and $F_{\rm noise}(t)$ is the total noise
with power density $S_{\rm total}^{\rm F}$.
The minimum force that drives the system into the nonlinear regime where bifurcation occur is $F_c$, with the corresponding
vibration frequency being $2\pi \omega_{c}$, and the amplitude $x_{c}$ can be calculated from Eq.~(\ref{Duffing})  \cite{Lifshitz2008}.
Near the nonlinear bifurcation point, we transform this equation to an over-damped one by following the standard procedure   \cite{Dykman1980PA}. For the case of modulation frequency $\Omega$ being much smaller than the decay rate $\omega_{0}/Q$ of the system, the signal to noise ratio (SNR)
is related to the system's noise power as  \cite{Luca1998RMP}
\begin{align}
\label{doublewellSNR}{\rm SNR}=\pi\frac{\gamma_k}{\delta \omega} x_m^2\left(\frac{\delta F_{\rm drive} m\omega}{S_{\rm total}^{\rm F}}\right)^2,
\end{align}
with $x_{m}$ the half of the vibration amplitude of the bistable states, $\delta \omega$ the nonlinearity frequency induced shift from linear resonance peak, and $\gamma_{k}$ the measured
random switching rate without modulation. So from the measured SNR we obtained $S_{\rm total}^{\rm F}$.
\end{methods}

\begin{addendum}
\item [Acknowledgements]
We thank L.\ Jiang for many stimulating discussions and comments and for his helps in improving the manuscript.
We thank Y.X.\ Liu, Z.Q.\ Yin, L.\ Tian C.P.\ Sun, and M.\ Lukin for helpful discussions. We thank Yiqun Wang from Suzhou Institute of
Nano-Tech and Nano-Bionics for nano-fabricating supports and
Guizhou Provincial Key laboratory of Computational Nano-Material Science for calculation supports.
This work was supported by the
973 Program (Grant No. 2013CB921800), the NNSFC (Grant Nos. 11227901,
91021005, 11104262, 31470835, 21233007, 21303175, 21322305, 11374305 and 11274299), the ``Strategic Priority Research
Program (B)'' of the CAS (Grant Nos. XDB01030400 and 01020000).

\item[Author contributions]
J.D. supervised the experiments. J.D. and P.H. proposed the idea and designed the experimental proposal. P.H. J.Z. L.Z. W.D. and F.X. prepared the experimental set-up.
P.H. J.Z. L.Z. and C.J. performed the experiments. J.Z. fabricated the sample. D.H. and X.Z. carried out the theoretical calculation.
M.C. carried out the finite elements simulation. P.H. C.D. F.X. and J.D. wrote the paper. All authors analyzed the data, discussed the results
and commented on the manuscript.

\item[Competing Interests] The authors declare that they have no competing financial interests.
\item[Additional information]
 Supplementary information accompanies the paper on http://www.nature.com/nnano.
Correspondence and requests for materials should be addressed to J.D.
\end{addendum}

\begin{figure}
\centering
\includegraphics[width=0.95\columnwidth]{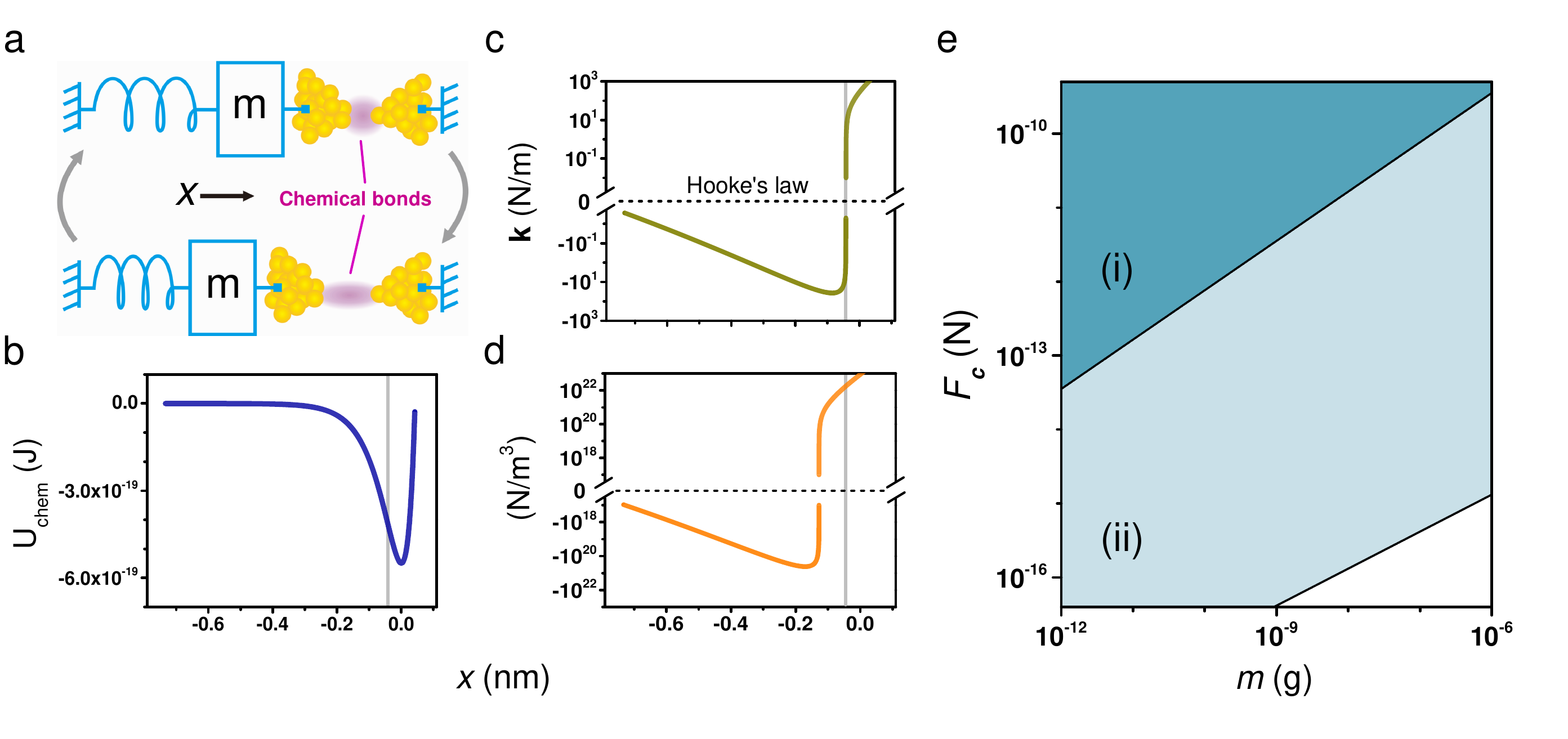}
    \caption{\textbf{Concept of the system and theoretical results.}
\textbf{a}, A macroscopic resonator tighten to an anchor via chemical bonds. The displaced resonator can compress (up) and stretch (down)
the chemical bond of gold-atom contact.
\textbf{b,c,d}, Density functional theory calculation results of (b) the chemical bonding interaction energy $U_{\rm chem}(x)$, (c) the modified  spring constant $\Delta k = -\partial^2 U_{\rm chem} / \partial x^2 $, and (d) the enhanced nonlinearity coefficient $\alpha = -(1/6)\partial ^4 U_{\rm chem}/ \partial x^4 $ as functions of the resonator displacement $x$.
\textbf{e}, Estimated threshold drive force $F_c$ as a function of the resonator mass, $m$, with: (i) intrinsic nonlinear response of the resonator (dark blue), (ii) enhanced nonlinear response of the resonator (light blue) by chemical bonding interaction, as indicated in gray line in (\textbf{b,c,d}).
where chemical bonding  induced linear response $\Delta k = 0$. (See Methods for details of the model.)
 }\label{dips}
\end{figure}

\begin{figure}
\centering
\includegraphics[width=0.95\columnwidth]{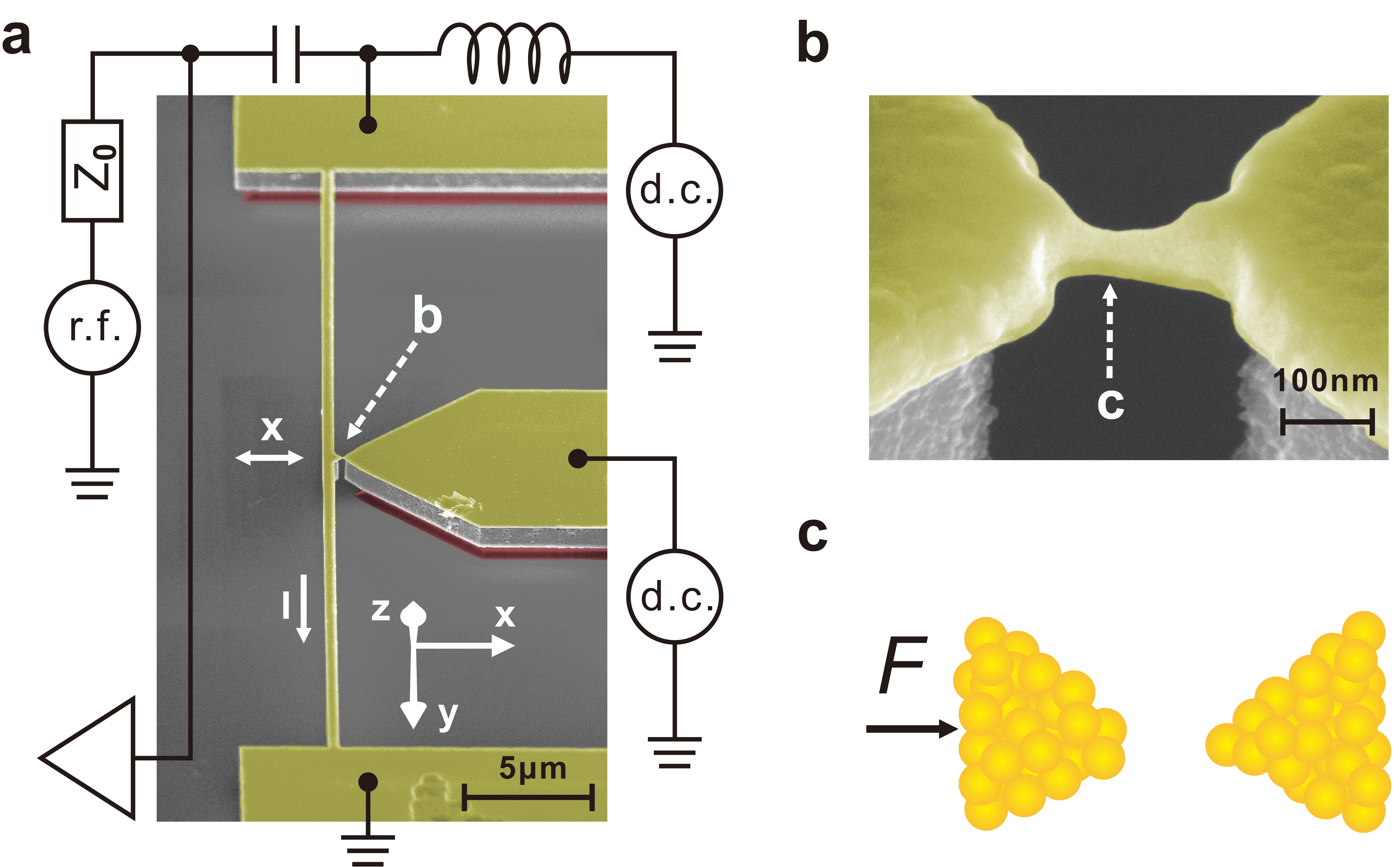}
    \caption{\textbf{Experimental setup.}
\textbf{a}, Scanning electron microscopy of the sample in false color.  The macroscopic resonator is a doubly clamped silicon beam
with thin layer of gold deposited on it, with dimension $l\times w \times t = 50~\mu$m $\times 1.5~\mu$m $\times 0.51~\mu$m and total mass approximately $0.2~{\rm ng}$. The center of the beam has horizontal displacement ($x$). In the presence of a $6~{\rm T}$ external magnetic field along the z direction, the electric current ($I$) can excite and detect the motion of the beam,
with the schematic circuits shown.
\textbf{b}, Nano-bridge connecting the beam to a stiff electrode, before experiments.
\textbf{c}, Cartoon plot of the atom contact generated on the nano-bridge indicated by `c' in \textbf{b}, the gold-gold bonding interaction is then tuned by
force $F$ which is controlled by a d.c. current through the beam, and the electrostatic interaction of the contact is minimized applying a d.c. bias on the tip.
 }\label{dips}
\end{figure}

\begin{figure}
\centering
\includegraphics[width=0.95\columnwidth]{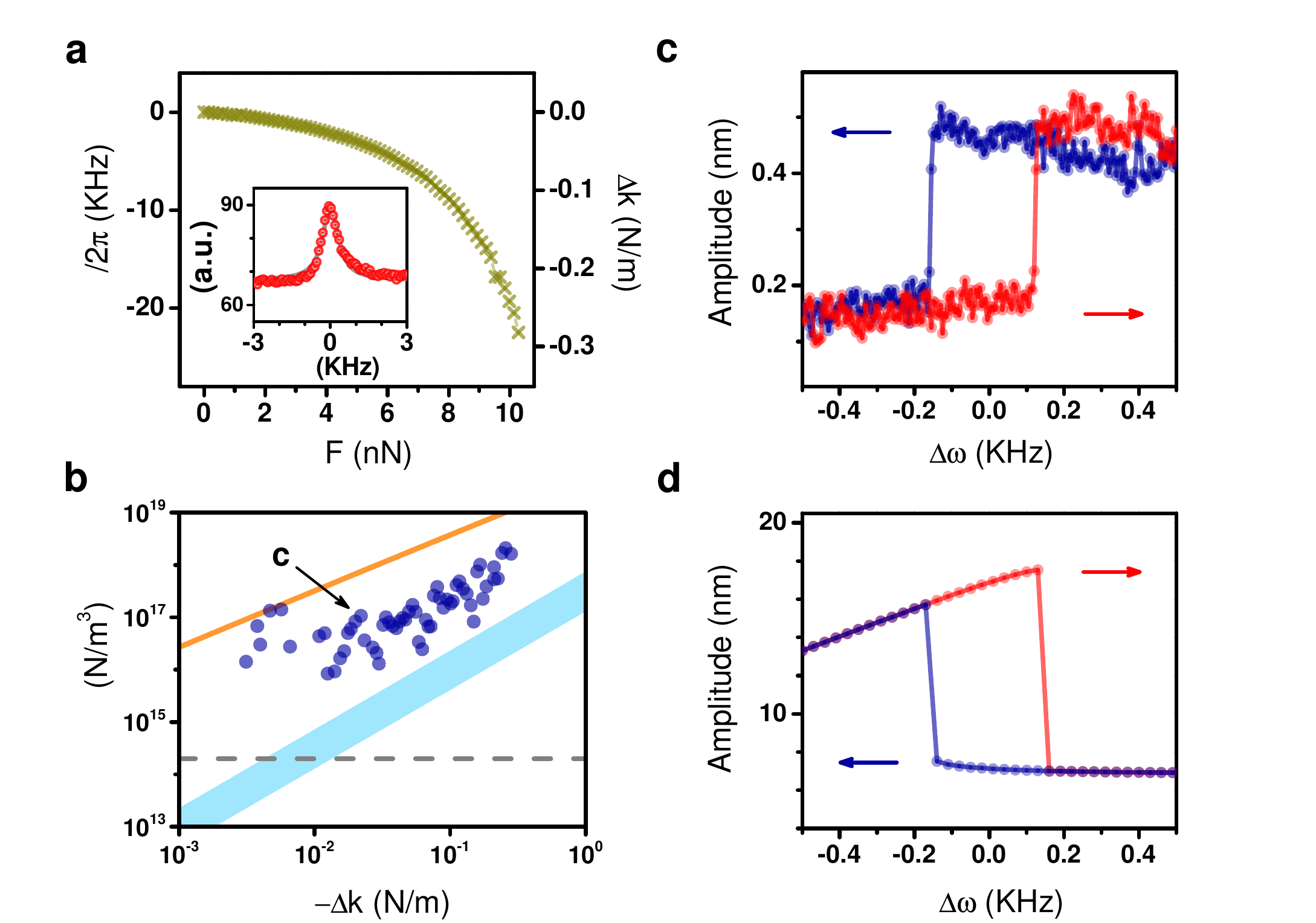}
\caption{\textbf{Tuning the macroscopic nonlinear response by chemical bonding force.}
\textbf{a}, Frequency shift (left axis) and the corresponding change in effective spring constant $\Delta k$ (right axis) as a
function of the external force $F$ that pushes the atom contact close to the beam, to do this,
frequency resonance under weak drive with the beam in near linear regime are measured (inset).
\textbf{b}, The tunable chemical bonding interaction can simultaneously change the Duffing nonlinear response $\alpha $ and the effective spring constant by $\Delta k$ (minus is for convenience). The measured Duffing constant (blue dots) is attributed to the short range chemical bonds (orange line), the long range van der Waals interaction (blue thick line), and the intrinsic Duffing constant due to the beam elongation (gray dashed line).
\textbf{c, d}, Hysteresis response of the system measured at the point indicated by 'c' in \textbf{b}, and measured with beam pulled far apart ($ \ge 30$ nm) from the stiff electrode so that only intrinsic nonlinearity due to the beam elongation contributes (\textbf{d}).
}
    \label{simulation}
\end{figure}

\begin{figure}
\centering
\includegraphics[width=0.65\columnwidth]{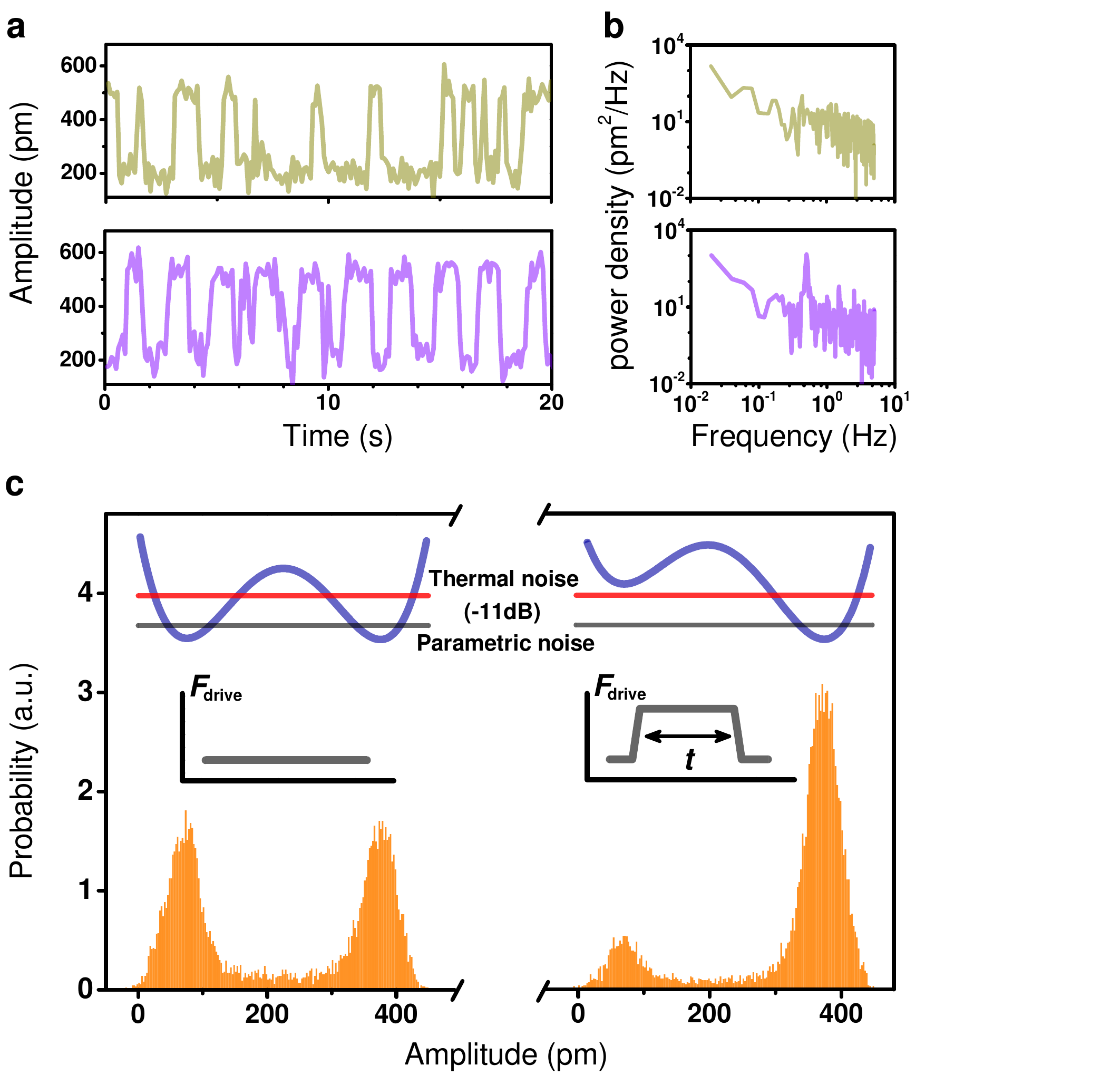}
    \caption{\textbf{Brownian thermal noise induced bistate transitions.}
\textbf{a},  Switching between bistable states with constant-amplitude drive (upper panel) and modulated-amplitude drive (lower panel). To observe bistate transitions, the driving frequency is fixed in the middle point of hysteresis response and the drive amplitude is $F_{\rm drive}=4 ~{\rm pN}$, and the modulation frequency and the amplitude are $\Omega = 0.5$ Hz and $\delta F_{\rm drive} = 0.18~{\rm pN}$, respectively.
\textbf{b},  Power spectrum density. Compared to the constant-amplitude drive (upper panel), the amplitude-modulated drive (lower panel) induces an additional peak in the power spectrum density at the modulation frequency $\Omega = 0.5$ Hz.
\textbf{c}, The amplitude distribution of the bistable resonator depends on the driving amplitude. Very different histograms for constant drive with $F_{\rm drive}=4.000 ~{\rm pN}$ (Left panel) and $F_{\rm drive}=4.002 ~{\rm pN}$ (Right panel) are observed. The change in amplitude distribution can be understood using an effective double well potential (sketched in the inset). The total electrical parametric noise is 11~dB below the Brownian thermal noise in our experiment.
 }
    \label{field}
\end{figure}


\begin{thebibliography}{99}
\bibitem{Aldridge2005PRL} Aldridge, J. S. \& Cleland, A. N. Noise-enabled precision measurements of a duffing nanomechanical resonator. \textit{Phys. Rev. Lett.} \textbf{94}, 156403 (2005).


\bibitem{Badzey2005N} Badzey, R. L. \& Mohanty, P. Coherent signal amplification in bistable nanomechanical oscillators by stochastic resonance. \textit{Nature} \textbf{437}, 995-998 (2005).

\bibitem{Stambaugh2006PRB} Stambaugh, C. \& Chan, H. B. Noise-activated switching in a driven nonlinear micromechanical oscillator. \textit{Phys. Rev. B} \textbf{73}, 172302 (2006).
\bibitem{Ono2008APE} Ono, T., Yoshida, Y., Jiang, Y. G., \& Esashi, M. Noise-enhanced sensing of light and magnetic force based on a nonlinear silicon microresonator. \textit{Appl. Phys. Express.} \textbf{1}, 123001 (2008).

\bibitem{Almog2007PRL} Almog, R., Zaitsev, S., Shtempluck, O. \& Buks, E. Noise squeezing in a nanomechanical duffing resonator. \textit{Phys. Rev. Lett.} \textbf{98}, 078103 (2007).

\bibitem{Venstra2013NC} Venstra, W. J., Westra, H. J. \& van der Zant, H. S. Stochastic switching of cantilever motion. \textit{Nature communications} \textbf{4}, 2624 (2013).





\bibitem{Karabalin2011PRL} Karabalin, R. B. \textit{\textit{et al.}} Signal amplification by sensitive control of bifurcation topology. \textit{Phys. Rev. Lett.} \textbf{106}, 094102 (2011).



\bibitem{Braakman2014APL} Braakman, F. R., Cadeddu, D., T\"{u}t\"{u}nc\"{u}oglu, G.,  Matteini, F.,  R\"{u}ffer, D., Fontcuberta i Morral, A., \& Poggio, M.,
 Nonlinear motion and mechanical mixing in as-grown GaAs nanowires. \textit{Appl. Phys. lett.} \textbf{105}, 173111 (2014).




\bibitem{Shim2007SCI} Shim, S. B., Imboden, M. \& Mohanty, P. Synchronized oscillation in coupled nanomechanical oscillators. \textit{Science} \textbf{316}, 5821 95-99 (2007)

\bibitem{Zhang2012PRL} Zhang, M. A. \textit{\textit{et al.}} Synchronization of micromechanical oscillators using light. \textit{Phys. Rev. Lett.} \textbf{109}, 233906 (2012).

\bibitem{Matheny2014PRL} Matheny, M. H. \textit{\textit{et al.}} Phase synchronization of two anharmonic nanomechanical oscillators. \textit{Phys. Rev. Lett.} \textbf{112}, 014101 (2014).
\bibitem{Mahboob2008NN} Mahboob, I. \& Yamaguchi, H. Bit storage and bit flip operations in an electromechanical oscillator. \textit{Nature Nanotech.} \textbf{3}, 275-279 (2008).

\bibitem{Venstra2010APL}Venstra, W. J., Westra, H. J. R. \& van der Zant, H. S. J. Mechanical stiffening, bistability, and bit operations in a microcantilever. \textit{Appl. Phys. Lett.} \textbf{97}, 193107 (2010)

\bibitem{Bagheri2011NN}Bagheri, M., Poot, M., Li, M., Pernice, W. P. H. \& Tang, H. X. Dynamic manipulation of nanomechanical resonators in the high-amplitude regime and non-volatile mechanical memory operation. \textit{Nature Nanotechnology} \text{6}, 726-732 (2011).


\bibitem{Eichler2011NN} Eichler, A. \textit{et al.} Nonlinear damping in mechanical resonators made from carbon nanotubes and graphene. \textit{Nature Nanotechnology} \textbf{6}, 339-342 (2011).

\bibitem{Chan2001PRL} Chan, H. B., Aksyuk, V. A., Kleiman, R. N., Bishop, D. J. \& Capasso, Federico.  Nonlinear Micromechanical Casimir Oscillator. \textit{Phys. Rev. Lett.} \textbf{87}, 211801 (2001).
\bibitem{Lee2002PRB} Lee, S. I., Howell, S. W., Raman,  A. \& Reifenberger, R.  Nonlinear dynamics of microcantilevers in tapping mode atomic force microscopy: A comparison between theory and experiment. \textit{Phys. Rev. B.} \textbf{66}, 115409 (2002).

\bibitem{Kozinsky2006APL} Kozinsky, I., Postma, H. W. Ch.,  Bargatin, I. \& Roukes, M. L.  Tuning nonlinearity, dynamic range, and frequency of nanomechanical resonators. \textit{Appl. Phys. Lett.} \textbf{88}, 253101 (2006).

\bibitem{Sankey2010NP} Sankey, J. C.,  Yang, C.,  Zwickl, B. M., Jayich, A. M.  \& Harris, J. G. E.  Strong and tunable nonlinear optomechanical coupling in a low-loss system. \textit{Nature Physics} \textbf{ 6}, 707-712 (2010).

\bibitem{Suh2010NL} Suh, J., LaHaye, M. D., Echternach, P. M., Schwab, K. C. \& Roukes, M. L. Parametric amplification and back-action noise squeezing by a qubit-coupled nanoresonator. \textit{Nano. Lett} \textbf{10}, 3990-3994 (2010).



\bibitem{Yurke1986PRL} Yurke, B. \& Stoler, D. Generating quantum-mechanical superpositions of macroscopically distinguishable states via amplitude dispersion. \textit{Phys. Rev. Lett.} \textbf{57}, 13 (1986).
\bibitem{Peano2004PRL} Peano, V. \& Thorwart, M. Macroscopic quantum effects in a strongly driven nanomechanical resonator, \textit{Phys. Rev. B} \textbf{70}, 235401 (2004).
\bibitem{Savel¡¯ev2006NJP} Savel'ev, S., Hu, X. D. \& Nori, F. Quantum electromechanics: qubits from buckling nanobars. \textit{New J. Phys.} \textbf{8}, 105 (2006).


\bibitem{Katz2007PRL} Katz, I., Retzker, A., Straub, R. \& Lifshitz, R. Signatures for a classical to quantum transition of a driven nonlinear nanomechanical resonator. \textit{Phys. Rev. Lett.} \textbf{99}, 040404 (2007).

\bibitem{Serban2007PRL} Serban, I. \& Wilhelm, F. K. Dynamical tunneling in macroscopic systems. \textit{Phys. Rev. Lett.} \textbf{99}, 137001 (2007).
\bibitem{Sillanpaa2011PRB} M. A. Sillanpaa, R. Khan, T. T. Heikkila, P. J. Hakonen, Macroscopic quantum tunneling in nanoelectromechanical systems, \textit{Phys. Rev. B} \textbf{84}, 195433 (2011).



\bibitem{Landau1986} Landau, L. D. \& Lifshitz, E. M. \textit{Theory of Elasticity} (Pergamon, Oxford, 1986).

\bibitem{Lifshitz2008} Lifshitz, R. \& Cross, M. C. \textit{Review of Nonlinear Dynamics and Complexity} (Wiley-VCH, 2009).


\bibitem{Dykman1980PA} Dykman, M. \& Krivoglaz, M. Fluctuations in nonlinear systems near bifurcations corresponding to the appearance of new stable states. \textit{Physica A: Statistical Mechanics and its Applications} \textbf{104}, 480-494 (1980).


\bibitem{PuHuang2013PRL} Huang, P. \textit{et al.} Demonstration of motion transduction based on parametrically coupled mechanical resonators. \textit{Phys. Rev. Lett.} \textbf{110}, 227202 (2013).

\setcounter{firstbib}{\value{enumiv}}
\end{thebibliography}

\begin{thebibliography}{9}
\setcounter{enumiv}{\value{firstbib}}
\bibitem{Giessibl2003RMP} Giessibl, F. J. Advances in atomic force microscopy. \textit{Rev. Mod. Phys.} \textbf{75}, 949-983 (2003).
\bibitem{Park1999APL} Park, H., Lim, A. K. L., Alivisatos, A. P., Park, J. \& McEuen, P. L. Fabrication of metallic electrodes with nanometer separation by electromigration. \textit{Appl. Phys. Lett.} \textbf{75}, 301-303 (1999).
\bibitem{Rubio-Bollinger2001PRL} Rubio-Bollinger, G., Bahn, S. R., Agrait, N., Jacobsen, K. W. \& Vieira, S. Mechanical properties and formation mechanisms of a wire of single gold atoms. \textit{Phys. Rev. Lett.} \textbf{87} 026101 (2001).

\bibitem{Cleland21999SAA} Cleland, A. \& Roukes, M. External control of dissipation in a nanometer-scale radiofrequency mechanical resonator. \textit{Sensors and Actuators A} \textbf{72}, 256-261 (1999).

\bibitem{Devoret2000NAT} Devoret, M. H. \& Schoelkopf, R. J. Amplifying quantum signals with the single-electron transistor. \textit{Nature} \textbf{406}, 1039-1046 (2000).

\bibitem{Clerk2010RMP} Clerk, A. A., Devoret, M. H., Girvin, S. M., Marquardt, F. \& Schoelkopf, R. J. Introduction to quantum noise, measurement, and amplification. \textit{Rev. Mod. Phys.} \textbf{82}, 1155-1208 (2010).
\bibitem{Luca1998RMP} Gammaitoni, L., H\"{a}nggi, P., Jung, P. \& Marchesoni, F. Stochastic resonance.\textit{Rev. Mod. Phys.} \textbf{70}, 223 (1998).

\end{thebibliography}
\end{document}